# Challenges and Opportunities of Computational Social Science for Official Statistics

*Serena Signorelli, Matteo Fontana, Lorenzo Gabrielli, Michele Vespe*
*European Commission, Joint Research Centre*

*Abstract*

The vast amount of data produced everyday (so-called 'digital traces') and available nowadays represent a gold mine for the social sciences, especially in a computational context, that allows to fully extract their informational and knowledge value. In the latest years, statistical offices have made efforts to profit from harnessing the potential offered by these new sources of data, with promising results. But how difficult is this integration process? What are the challenges that statistical offices would likely face to profit from new data sources and analytical methods? This chapter will start by setting the scene of the current official statistics system, with a focus on its fundamental principles and dimensions relevant to the use of non-traditional data. It will then present some experiments and proofs of concept in the context of data innovation for official statistics, followed by a discussion on prospective challenges related to sustainable data access, new technical and methodological approaches and effective use of new sources of data.

*Introduction*

Official statistics can be defined as the ensemble of all indicators, statistics and indices that are produced and disseminated by national statistical authorities (OECD et al., 2002). Right now, in their operations, official statistics tend to rely on so-called *traditional data sources*, namely *census data*, *surveys* and *administrative data*[1].

Yet, in an era characterised by increasing amounts of time spent living with connected devices, large amounts of new data are generated and collected every day. The places that we live in or that we visit can be inferred by analysing the position marked by our smartphones, our passions and relationship networks inferred from what we write on social media, our health status from physiological data gathered through smart watches.

By living in a world that is a hybrid between its real and virtual instances, every day we leave traces and footprints of our life that are digital and can thus be collected, stored and processed.

Is it possible for statistical offices to draw on these "digital trace" data for creating new statistical indicators or for improving speed, quality and resolution of old ones in the field of social sciences? Such questions are very timely and high policy relevance, as shown by the collective exercise carried out at the Joint Research Centre of the European Commission (2022) with the aim of mapping the demand side of Computational Social Science for Policy and its specific chapter on data innovation for official statistics. In this chapter we will address main challenges and needs that statistical authorities will have to face in order to harness the full potential of these new data sources and illustrate some successful examples, with a focus on Computational Social Science.

---

[1] Examples include birth and death registers in demographic statistics or the registries of real estate transactions in housing market statistics.

*Current Official Statistics systems*

In a recent report (2019, Chapter 7), The United Nations define three different types of data sources that are or could be used in official statistics:

1. Statistical data sources, composed by data collections created primarily for statistical purposes. This category includes surveys and census data.
2. Administrative data sources, which, differently from the former sources, are primarily set up for administrative purposes by public sector bodies.
3. Other data sources represented by all other sources created for commercial, market research or other private purposes.

The third source of data is the one that usually is referred to as the term "big data".

In the following section we introduce the official statistics principles and how these new data sources relate to them. The section provides an overview of the steps that statistical agencies have undertaken so far to discover their potential and leverage their value and represent the foundations for Computational Social Science to provide input to policy through Official Statistics.

*Statistical principles*

To fulfil their mission of providing timely and reliable data, the National Statistical Systems must comply with a set of principles that were formalised and adopted for the first time in 1991 by the Conference of European Statisticians (1991), revised afterwards and adopted globally by the UN Statistical Commission (1994)[2], with the name of Fundamental Principles of Official Statistics. Subsequently these principles have been updated periodically: the most recent version dates to 2013 (UN Economic and Social Council, 2013).

Together with the principles above, the concept of "quality" of official statistics needs to be taken into account. Brackstone (1999) defined quality in statistical agencies as "embracing those aspects of the statistical outputs of a NSO [National Statistical Office] that reflect their fitness for use by clients" and, as this concept is not capable of giving an operational definition, defined six dimensions of the broader concept to quality (see **Table 1**).

**Table 1: The Six Dimensions of Data Quality, from** *Brackstone (1999)*

---

[2] The United Nations Statistical Commission represents he highest body of the global statistical system and brings together the Chief Statisticians from member states from around the world

| Relevance | "The **relevance** of statistical information reflects the degree to which it meets the real needs of clients. It is concerned with whether the available information sheds light on the issues of most importance to users. Assessing relevance is a subjective matter dependent upon the varying needs of users. The NSO's challenge is to weigh and balance the conflicting needs of different users to produce a program that goes as far as possible in satisfying the most important needs and users within given resource constraints." |
|---|---|
| Accuracy | "The **accuracy** of statistical information is the degree to which the information correctly describes the phenomena it was designed to measure. It is usually characterized in terms of error in statistical estimates and is traditionally decomposed into bias (systematic error) and variance (random error) components. It may also be described in terms of the major sources of error that potentially cause inaccuracy (e.g., coverage, sampling, nonresponse, response)." |
| Timeliness | "The **timeliness** of statistical information refers to the delay between the reference point (or the end of the reference period) to which the information pertains, and the date on which the information becomes available. It is typically involved in a trade-off against accuracy. The timeliness of information will influence its relevance." |
| Accessibility | "The **accessibility** of statistical information refers to the ease with which it can be obtained from the NSO. This includes the ease with which the existence of information can be ascertained, as well as the suitability of the form or medium through which the information can be accessed. The cost of the information may also be an aspect of accessibility for some users." |
| Interpretability | "The **interpretability** of statistical information reflects the availability of the supplementary information and metadata necessary to interpret and utilize it appropriately. This information normally covers the underlying concepts, variables and classifications used, the methodology of collection, and indications of the accuracy of the statistical information." |
| Coherence | "The **coherence** of statistical information reflects the degree to which it can be successfully brought together with other statistical information within a broad analytic framework and over time. The use of standard concepts, classifications and target populations promotes coherence, as does the use of common methodology across surveys. Coherence does not necessarily imply full numerical consistency." |

These six dimensions have been adapted by the main International Statistical Organizations to their own needs, as detailed in this table published by UNECE (Vale, 2010):

**Table 2: Mapping Quality Components Used by International Statistical Organisations, from** *Vale (2010)*

| UNECE | OECD | EUROSTAT | IMF |
|---|---|---|---|
| Relevance | Relevance | Relevance | Pre-requisites of quality (part) |
| | | | Methodological soundness |
| Accuracy | Accuracy | Accuracy | Accuracy and |

|                                              |              |                          | Reliability                     |
| -------------------------------------------- | ------------ | ------------------------ | ------------------------------- |
| Timeliness                                   | Timeliness   | Timeliness and           | Serviceability (part)           |
| Punctuality                                  |              | Punctuality              |                                 |
| Accessibility                                | Accessibility | Accessibility and       | Accessibility                   |
| Clarity                                      | Interpretability | Clarity              | Assurances of integrity (part)  |
| Comparability                                | Coherence    | Comparability            | Serviceability (part)           |
|                                              |              | Coherence                |                                 |
|                                              | Credibility  |                          |                                 |
| (Considered more relevant at the level of the organisation) |  |          | Pre-requisites of quality (part) |
|                                              |              |                          | Assurances of integrity (part)  |

When dealing with new data sources, one of the key elements to be considered is **timeliness**. Surveys and censuses usually require a substantial timeframe between the collection phase and the publication of results, while different sources like, for example, mobile phone data, could be available, at least theoretically, in near-real time. Together with timeliness, this highlights an additional feature offered by new data sources that is the potential to improve **frequency** or periodicity of the data collected. The time between observation can be reduced almost arbitrarily below the yearly or monthly that are typical in current official statistics.

As Brackstone (1999) points out in **Table 1**, "Timeliness is typically involved in a trade-off against accuracy". In fact, this is specifically true for traditional data sources such as survey or census data. With reference to new sources of data, they usually do not constitute a representative sample of the population marking an intrinsic limitation to accuracy. In the case of new data sources, accuracy is less linked to timeliness given the availability of information that occurs in almost real-time. When dealing with innovative data sources, other kind of trade-offs may emerge; as an example, when dealing with mobility data gathered via mobile networks (as done in Iacus et al. (2020), accuracy could be in trade-off with resolution, since the increase in granularity may further reduce the representativeness of the information.

This representation issue constitutes one of the differences between data coming from research institutions and from commercial companies highlighted by Liu et al. (2016). Private companies do not necessarily follow scientific data collections procedures or statistical sampling schemes, as their main objective is to streamline processes such as billing (e.g., Call Detail Records – CDR – from Mobile Network Operators), or optimise services as product recommendations and advertising (e.g., social media advertising platform data), ultimately maximising their profit. Another accuracy aspect that Liu et al. (2016) highlight is the fact that private companies could "change the sampling methods and processing algorithms at any time and without any notice", adding uncertainty and risk to accuracy. Examples of this were reported when accessing mobility data from multiple Mobile Network Operators in Europe to help fight COVID-19 (Vespe et al., 2021). Finally, Liu et al. (2016) emphasize how the validity of data itself could be at risk, as "commercial platforms have no obligation or motivation to ensure the authenticity and validity of the data they collected".

The dimensions described above are not the only ones affected by the uptake of innovative data sources in official statistics; we need to consider **accessibility** issues, as new data sources – often privately-held - may be difficult or expensive to procure. At the same time, it is also true that the data sources currently used in official statistics (surveys, generally) already present an increase in nonresponse rates, that leads to a reduction in the quality of data and consequently to an increase in the associated costs (Luiten et al., 2020). In order to improve accessibility, a switch to new data sources could be framed as a possible way to address rising costs associated with traditional data collections. Costs would probably not be reduced, but financial resources could be invested into new sources that could complement (or even replace) existing ones. Nevertheless, in many other cases, data may not be yet available on the market for several reasons (e.g., non-clear reputational or monetisation advantages over risks of non-compliance after sharing the data), requiring additional efforts to improve such data flows, including regulatory ones (e.g., the EU Data Governance Act[3] or the EU Data Act[4]).

As mentioned, the use of such new data sources for official statistics would be a *secondary* one with respect to the reasons for which they were conceived and collected. For example, CDR data could be employed for mobility analysis (Blondel et al., 2015), while social media advertising platform data could be used to estimate population flows (Spyratos et al., 2019). This requires a certain amount of additional processing and interpretation in order to lead to meaningful indicators. **Interpretability**[5] will therefore play a significant role in the future of official statistics with new data, as it will not be straightforward as currently is with surveys and census data, designed and set up to describe the phenomenon they are supposed to measure.

Also, **coherence** will be affected, as it will be important for these data sources to be sustainable over time, making them available continuously and with constant underlying methodology (this links again to accuracy), or at least with full knowledge of it to be constantly updated as part of the production process.

Going back to the Fundamental Principles of Official Statistics (UN Economic and Social Council, 2013), which deal with issues like accountability, relevance, impartiality and transparency among others, it can be observed that a process of adaptation of these guidelines to a new paradigm will be needed. For example, principle no. 2 states that "*to retain trust in official statistics, the statistical agencies need to decide according to strictly professional considerations, including scientific principles and professional ethics, on the methods and procedures for the collection, processing, storage and presentation of statistical data*". This principle refers to transparency in official statistics, which is assessed and guaranteed by a set of guidelines that must be fulfilled by professionals handling data in statistical offices.

---

[3] https://eur-lex.europa.eu/legal-content/EN/TXT/?uri=CELEX%3A52020PC0767
[4] https://digital-strategy.ec.europa.eu/en/library/data-act-proposal-regulation-harmonised-rules-fair-access-and-use-data
[5] Interpretabiliy here has to be intended in the broader sense used by Brackstone (1999) and not as in the machine learning context (see for example Murdoch et al. (2019)), where it is more related to algorithmic transparency.

Nevertheless, the concept of transparency applied to new sources and digital trace data should not only be seen from a 'data handler' perspective, but it must be complemented by a set of rules that refer to procedures and codes used to produce insights, calling for open-source practices and FAIR (Findable, Accessible, Interpretable and Reusable) data principles (Wilkinson et al., 2016) to ensure interpretability and reproducibility.

The structure of the statistical system may need to adapt when using digital trace data: the "survey design" part would become less relevant in this context, possibly superseded by a "data ingestion" and "data processing" sections, while processing becomes central.

The nature and composition of the tasks a NSO needs to perform to deliver reliable official statistics starting from big data may call for an adaptation of the organisational structure as well as of the competences needed by NSOs.

Many statistical offices have begun this transformation with exploratory exercises with the exception of sporadic cases[6]. This is a challenge that statistical offices may need to face. As an example, in terms of computer code, with a one-off analysis (as done with scientific research) it is sufficient to publish the code as open source, while in regular production settings of official statistics, the code itself needs to be maintained, implementing regular edits and versioning. This translation of statistical methodology into software code has been introduced by Ricciato (2022) with the name of *softwarization of statistical methodologies*.

*Recognition of the Value of New Data Sources*

In Europe, the European Statistical System[7] (ESS) has been involved in recognising the existence of digital trace data and its value since nearly a decade.

Two documents have paved the way to the use of innovative data sources in official statistics. The *Scheveningen Memorandum - Big Data and Official Statistics* (DGINS, 2013), represents the first statement through which the ESS recognised the importance of these new data sources and highlighted the main issues related to their use.

The *Bucharest Memorandum on Official Statistics in a Datafied Society (Trusted Smart Statistics)* (DGINS, 2018), represents an updated version of the former document, where the ESS underlines the need for *"amendments to the statistical business architecture, processes, production models, IT infrastructures, methodological and quality frameworks, and the corresponding governance structures"*.

Moreover, in 2021 Eurostat[8] started a revision process of Regulation 223/2009[9] (the EU legal framework for European statistics) considering the new needs of official statistics. The updated

---

[6] International travel statistics in Estonia: https://statistika.eestipank.ee/failid/mbo/valisreisid_eng.html and foreign visitor statistics in Indonesia: https://www.bps.go.id/subject/16/pariwisata.html#subjekViewTab1, both using mobile positioning data

[7] The partnership between the European Community statistical authority, composed by Eurostat, the national statistical offices (NSOs) and other national authorities in each EU Member State that are responsible for the development, production and dissemination of European statistics

version of the Regulation is expected to be finalised by the end of 2022. One of the explicit goals of the revision process is to set the legal framework for the reuse of privately held data for the development, production, and dissemination of official statistics in Europe (Baldacci et al., 2021).

On a more international perspective, the Organisation for Economic Co-operation and Development (OECD) collected a series of examples of statistical applications (OECD, 2015) that made use of new data sources, as well as a list of limitations of this type of data. More importantly, the report introduces the implications for statistical offices when using these new data sources. Specifically, they envision three different possible roles for statistical offices, which may:

1. act as certificatory institutions (giving the so-called 'trust mark' to datasets);
2. act as dissemination institutions (in a way that all statistics produced with non-traditional data are stored and disseminated in a central agency – the National Statistical Office);
3. become active users of non-traditional data sources, for complementing the traditional ones as well as to create standalone statistical series.

The OECD has already underlined some sensitive issues that will need to be taken into consideration for a successful adoption of non-traditional data in the workflow of national statistical offices. The main challenges are represented by the acquisition of skills needed to work with non-traditional data, the relevant data governance principles as well as privacy concerns (OECD, 2015). They also see space for partnerships of National Statistical Offices with universities and research organisations to best exploit the new opportunities brought by data innovation and to become collectors and disseminators of best practices.

*Some Proof of Concepts and Experiences*

In 2014, the United Nations established a Global Working Group (GWG) on Big Data for Official Statistics[10] with the aim of promoting the practical use of big data sources, as well as building trust in the use of these sources for official statistics.

One of the outputs of the group was a handbook on the use of mobile phone data for official statistics (UN Global Working Group on Big Data for Official Statistics, 2019), which put forward a series of practical examples of the use of this data source in different statistical domains (tourism, population, migration, commuting, traffic flow and employment). Many countries (Estonia, Japan, Sri Lanka among others) launched pilots and projects that have some potential for statistics in the mentioned statistical domains.

Most practical examples of applications have been carried out by European countries, where a partnership between the Eurostat, the NSOs and other National authorities that are responsible for the development, production and dissemination of European statistics was implemented with the name of European Statistical System (ESS).

---

[8] The statistical office of the European Union
[9] https://eur-lex.europa.eu/legal-content/EN/ALL/?uri=CELEX%3A32009R0223
[10] https://unstats.un.org/bigdata/

One of the first attempts identified is ESSnet Big Data I[11], composed by 22 NSOs. The objective of this initiative is to integrate big data into the regular production of official statistics. This is achieved via the development of projects that could explore the potential of these data sources, carried out from February 2016 to May 2018.

One of these projects was carried out with the help of six national statistical institutes (and afterwards other four joined) and investigated the feasibility of using job advertisement data scraped from the Web to improve official estimates of job vacancy statistics[12]. The activity consisted in the comparison between online job advertisement and job vacancy surveys. Some cases demonstrated a high correlation, while others showed only a loose relationship between the two. Nevertheless, this appears to be a promising area where innovative data can complement traditional survey data by potentially producing flash estimates or increasing the frequency survey-based statistics, but also to produce additional insights about occupations, required skills and labour demand in local areas.

Another ESSnet example aimed at inferring enterprise characteristics by accessing their websites through Web scraping techniques[13]. Six NSOs were involved, and their activity focused on six different use cases (URLs retrieval, e-commerce/web sales, social media detection, job advertisement detection, NACE[14] detection, SDGs detection) using both deterministic and machine learning methods. The predicted values can be used at unit level, to enrich the information contained in the register of the population of interest, and at population level, to produce estimates. The activity resulted in a series of output indicators, published as experimental statistics)[15].

A third example in the framework of the ESS network (European Statistical System, 2017) concerned the use of scanner data or web-scraping for Consumer Price Index (made by NSOs in France, Italy, the Netherlands, Poland and Portugal), the use of mobile phone data to study population and the study of tourist accommodations offered by individuals (French NSO), an analysis on the identification of inhabited addresses through electricity providers data to reduce survey costs (NSOs in Poland and Estonia), the use of credit and debit cards data in the National Accounts (Portuguese NSO).

A deeper analysis was carried out specifically on tourism statistics. Eurostat has made an extended analysis of data sources having potential relevance for measuring tourism. In a recent report (2017), Demunter develops a taxonomy of big data sources relevant to tourism, including: communication systems (e.g., MNO data, social media posts), Web (e.g., web activity data), business process generated data (e.g., flight bookings, financial transactions), sensors (e.g., earth observation, vessel tracking systems, smart energy meters) and crowd sourcing (e.g., Wikipedia, OpenStreetMap).

---

11 https://ec.europa.eu/eurostat/cros/essnet-big-data-1_en
12 https://ec.europa.eu/eurostat/cros/content/wp1-reports-milestones-and-deliverables1_en
13 https://ec.europa.eu/eurostat/cros/content/wp2-reports-milestones-and-deliverables1_en
[14] NACE stands for the statistical classification of economic activities in the European Community
[15] https://ec.europa.eu/eurostat/web/experimental-statistics/

An attempt to develop a hybrid between one-off analyses and regular production statistics has been undertaken by some statistical offices in the form of experimental statistics. Among the examples that can be identified, a very notable one was carried out by Eurostat[16]. These statistics cover 14 topics[17], ranging from collaborative economy platforms to skills mismatch. All these experiments are listed and can be further explored[18]. They are deemed *experimental* as they "have not reached full maturity in terms of harmonisation, coverage or methodology". Nevertheless, the potential in terms of provided insights and knowledge of such solutions is clearly disruptive. Moreover, in a spirit of experimentation and co-creation, Eurostat and the single NSOs invite users to submit feedback and suggestions to improve them.

The UK Office for National Statistics published on its website a guide on experimental statistics[19], defining the features of this kind of statistics, namely:

- "new methods, which are being tested and still subject to modification;
- partial coverage (for example, of industries) at that stage of the development programme;
- potential modification following user feedback about their usefulness and credibility compared with other available statistical sources".

*The Need for Change*

The above considerations and examples show the significant attention by statistical offices on the use of novel data sources since almost a decade, as well as the readiness and will to innovate. But what does this shift mean in practice for them?

With the availability of new data sources, the statistical system may need to adapt, as it was traditionally designed to work with data of a different nature (surveys and administrative data). This comes from the fact that data from new sources (that we will call *non-traditional data* for convenience), are quite different from traditional ones:

- Firstly, surveys are designed to produce specific statistics, whereas non-traditional data are collected for other purposes (see Section *Current Official Statistics systems*).

- Secondly, while non-traditional data tend to capture human behaviour, they are not directly generated by humans, but by automated systems and machines with which humans interact. These peculiarities require an additional effort in terms of translating machine logs into information about human actions, and then connecting such actions to human behaviour.

---

[16] https://ec.europa.eu/eurostat/web/experimental-statistics
[17] At the the moment of publishing
[18] https://ec.europa.eu/eurostat/web/experimental-statistics/overview/ess)
[19] https://www.ons.gov.uk/methodology/methodologytopicsandstatisticalconcepts/guidetoexperimentalstatistics

- Thirdly, the process of translation of machine logs into information requires many choices to be done by researchers, that will in some way influence the result (Ricciato, 2022; Ricciato et al., 2021).Having understood the context to properly address these issues, we can observe how the analysis and use of new data for official statistics requires a dual set of competences: both in terms of modelling and inference of human behaviour (in its many possible dimensions) as well as the technical capabilities needed to manage and analyse such big and complex data sources. This specific skillset is the one required by the emerging field of Computational Social Science. Statistical authorities may need to develop and strengthen these skills to benefit from the information included into non-traditional data.

These new data sources represent a huge opportunity for statistical offices to innovate while increasing openness. Nonetheless, challenges relevant to data access, adaptation of processes and effective uses of the data will have to be addressed.

*Data Access*

The great majority of the data sources that could be harnessed for official statistics purposes reside with the private sector. The debate on the access to such data is broad and vivid, with different opinions arising, in favour and against the mandatory obligation for private companies of giving access to the data.

The European Commission is addressing this issue in its legislative process and has recently proposed a regulation in the framework of the European Data Strategy, the Data Act[20] that, among other provisions, aims at fostering business-to-government data sharing for the public interest, supporting business-to-business data sharing, and evaluating the Intellectual Property Rights (IPR) framework with a view to further enhance data access and use. The legislative process started in May 2021 and included a public consultation carried out during summer of 2021 that led many affected parties to the publication of a number of position papers. From the perspective of statistical offices, the ESS called on the need for the Data Act to ensure that European Statistical Offices and Eurostat can be granted access to privately held data for the development, production and dissemination of official statistics (European Statistical System, 2021). On the other hand, private sector data holders stressed on a lack of incentives to share data and an unclear impact that this sharing would have in practice (Bitkom, 2021), but also on voluntary sharing of data (and not an obligation) (AmCham EU, 2021; ETNO, 2021; Orgalim, 2021) as well as legitimate business interest around data to be protected. The Data Act was proposed by the Commission on 23 February 2022 (COM, 2022), providing means for public sector bodies, EU institutions, agencies or bodies to access and use privately held data in exceptional circumstances such as in emergencies. Such data may be shared to carry out scientific research activities compatible with the purpose for which the data was requested by the public sector body, or with national statistical institutes for the compilation of official statistics.

---

[20] https://digital-strategy.ec.europa.eu/en/library/data-act-proposal-regulation-harmonised-rules-fair-access-and-use-data

Guidelines and best practices are also being published in the literature, such as by researchers from Bank of Italy, highlighting the three main challenges that characterise the access and use of new data sources: trust, usability and sustainability (Biancotti et al., 2021). Moreover, the authors developed a set of principles that should guide data partnerships and that concern general aspects, principles specifically directed to statistical agencies and to private-sectors data collectors (Biancotti et al., 2021). The principles directly related to statistical offices build around three main notions: responsibility and accountability (on process, output and methodology), safeguard (of individual and business interests), coordination and standardisation (the "collect only once" principle, to avoid the same request to the same data provider).

*Adapting the Official Statistics System*

In a recent paper (2020) Ricciato and co-authors highlight a set of important challenges that statistical offices may need to address when confronted with the possibility of using non-traditional data, which imply a series of changes "[…] in almost every aspect of the statistical system: processing methodologies, computation paradigms, data access models, regulations, organizational aspects, communication and disseminations approaches, and so forth".

Going more into practical details and on specific issues, one of the most critical is **privacy** that must be protected via e.g. Privacy Enhancing Technologies (PET).

Borrowing greatly from the work of Ricciato and co-authors (2019), the UN Big Data Working Group defined in 2019 the three goals that need to be taken as guidelines when dealing with privacy concerns: input privacy, output privacy and policy enforcement (Big Data UN Global Working Group, 2019). In particular, "one or more Input Parties provide sensitive data to one or more Computing Parties who statistically analyse it, producing results for one or more Result Parties" (Big Data UN Global Working Group, 2019). The first goal, *input privacy*, must ensure that Computing parties are not able to access (or to indirectly derive with specific techniques and mechanism) any input value provided by Input Parties. At the other end of the process, *output privacy* has to ensure that published results do not contain identifiable input data. The third goal - introduced by the Big Data UN Global Working Group (2019), *policy enforcement*, represents the meeting point of the first two, as it is able to assure that they are automatically assured in a privacy-preserving statistical analysis system. Without entering into many details, this goal is concretised if there exists a mechanism that allows Input Parties to exercise positive control over computations that can be performed on sensitive inputs and over the publication of results; the just mentioned positive control is "[…] expressed in a formal language that identifies participants and the rules by which they participate" and carried out through a series of rules and decision points.

The report then presents five different PETs for statistics: Secure Multiparty Computation, (Fully) Homomorphic Encryption, Trusted Execution Environments, Differential Privacy and Zero Knowledge Proofs. In light of the above-mentioned system, for each PET they describe which of the three goals it supports and in which way.

Another important issue is the **transparency** of National Statistical Offices. Luhmann et al. (2019) propose a new paradigm called STATPRO (Shared, Transparent, Auditable, Trusted, Participative, Reproducible and Open). The authors make an open call to all National Statistical Offices about the need of implementing these seven principles, in order to achieve the goal of having a transparent and defensible evidence-based data-informed policymaking. In particular, some best practices from the Open-Source Software (OSS) community are needed for the development and deployment of statistical processes. As an example, they suggest that algorithms and methods should be available and accessible to anyone, with adequate level of documentation, and versioning should be introduced for environments.

After looking at specific issues, we need to focus on how to practically adapt the production system with the new requirements brought by the use of non-traditional data. One possible approach is proposed by Grazzini et al., (2018) through the so-called *plug and play* design. This approach was thought to handle the changes needed in production systems, and it is based on software components, which are modular and customisable, that are subsequently assembled together. This design has the advantage of allowing the integration of existing systems, operations and components with the new ones needed to embrace new data and/or models. Being modular, it also allows to overcome the constraint usually present on the choice of platform used for the implementation.

One practical proposal that has been theorised and discussed in Europe in recent years is the introduction of Trusted Smart Statistics. One of the main principles behind this proposal relies on the idea of "pushing computation out instead of pulling data in" (Ricciato, Wirthmann, et al., 2019). The concept is often referred to as "in situ data processing" (Höcük, 2021). This implies that the new data sources that statistical offices wish to analyse and integrate with traditional ones do not necessarily need to leave the premises of the data holders. Instead, the algorithm will reach the latter in order to perform computations, and afterwards only aggregated and processed data will be led to statistical offices to produce official statistics. On the one hand, this new paradigm will allow to preserve the privacy principle (as the data are not leaving their premises), but on the other hand more attention must be paid to transparency and accountability. One way to address these issues is the way already paved by the OPen ALgorithm project (OPAL), which declared algorithmic transparency as its foundational principle[21]. The proposal consists in making open by default all the software code along the whole data processing chain, and allowing everybody to see it and, eventually, audit it (Ricciato, Wirthmann, et al., 2019).

*Effective Use of the New Sources*

Once the first two issues are addressed (access to the data and changes in the statistical system), an important one (if not the most important) remains: what are the new statistical products that could only be developed using these new data sources? And why would be the responsibility of statistical offices to take care of this (and not, for example, a local authority)?

---

[21] http://www.opalproject.org/about-opal

This implies that the focus must now go to the demand side, and to the identification of the questions that statistical offices could tackle with these data sources, in line with what proposed by the European Commission, Joint Research Centre (2022) in the Computational Social Science for Policy mapping exercise. This represents a challenging task, as statistical offices need to take some time and reflect on what to highlight, but also why this relies in their mandate, and not among some other institution's activity.

Some examples of these new 'needs' are clearly shown for instance in Romanillos Arroyo & Moya-Gómez (2022), Napierala & Kvetan (2022), Manzan (2022) and Crato (2022). Concerning tourism, for example, after an introduction about new data sources and new computational methods for the tourism sector, the authors propose a series of potential applications (in the form of KPIs) on environmental impact and socio-economic resilience of tourism. By looking at the KPIs proposed to monitor land use related to the tourism activities, for instance, one of the indicators put forward aims at quantifying the presence of short-term rentals platforms (like Airbnb) through the analysis of accommodation platform data or similar. This indicator would allow to get more insights about a phenomenon that is increasing and that is not captured through traditional data sources in the tourism sector (namely, surveys) (Romanillos Arroyo & Moya-Gómez, 2022).[22]

Another example concerns direct and indirect water consumption at tourism destinations, a KPI that could be useful for the management of resources consumption related to leisure places. In this case different datasets could be used: from smart meters to food consumption data that in turn can be inferred from credit card data (Romanillos Arroyo & Moya-Gómez, 2022). As can be seen, these new proposed indicators require prior agreement to accessing the data, and therefore the three issues we presented in this chapter again show their very close connectedness.

*Discussion and Way Forward*

Summarising the issues highlighted in this chapter on the use of Computational Social Science for official statistics, the focus goes to the three main enablers:

- the access to the data
- the adaptation needed by official statistical system, and
- new statistical products that could be developed using these new data sources.

Concerning this last point, a proposal to facilitate the implementation of this could be the institution of specific committees or steering groups with the aim to discuss possible solutions to the issues presented. Something that needs to be underlined is the fact that even if data access could come for free (following specific partnerships or law provisions), the processing of these new data sources has a cost.

---

[22] The phenomenon of short-term rental accommodation in tourism is already under the lens of the European Commission, that will shortly propose a regulation about it (https://ec.europa.eu/info/law/better-regulation/have-your-say/initiatives/13108-Tourist-services-short-term-rental-initiative/public-consultation_en)

As a concluding remark, these new data sources have enormous potential for the official statistics world in terms of improved timeliness and granularity, but they can only be considered as a complementary source, and not pure substitutes of the traditional ones. As it is thoroughly explained in this chapter, due to the strict statistical requirements in terms of quality of the data used in official statistics we think that these new sources of data could improve and complement the existing ones.